\documentclass[conference]{IEEEtran}
\IEEEoverridecommandlockouts
\usepackage{cite}
\usepackage{amsmath,amssymb,amsfonts}
\usepackage{algorithmic}
\usepackage{graphicx}
\usepackage{textcomp}
\usepackage{xcolor}
\def\BibTeX{{\rm B\kern-.05em{\sc i\kern-.025em b}\kern-.08em
    T\kern-.1667em\lower.7ex\hbox{E}\kern-.125emX}}
\begin{document}

\title{CNN-DRL with Shuffled Features in Finance\\
{\footnotesize Short paper (CSCI-RTAI)}
}

\author{
\IEEEauthorblockN{1\textsuperscript{nd} Sina Montazeri}
\IEEEauthorblockA{\textit{CSCE Department} \\
\textit{University of North Texas}\\
Denton, TX, USA  \\
SinaMontazeri@my.unt.edu}
\and
\IEEEauthorblockN{2\textsuperscript{rd} Akram Mirzaeinia}
\IEEEauthorblockA{\textit{Independent Researcher} \\
\textit{Not Affiliated }\\
Ajman, United Arab Emirates\\
mirzaei.kad@gmail.com}
\and
\IEEEauthorblockN{3\textsuperscript{th} Amir Mirzaeinia}
\IEEEauthorblockA{\textit{CSCE Department} \\
\textit{University of North Texas}\\
Denton, TX, USA \\
amir.mirzaeinia@unt.edu}
\and
% \and
% \and
}

\maketitle

\begin{abstract}
In prior methods, it was observed that the application of Convolutional Neural Networks agent in Deep Reinforcement Learning to financial data resulted in an enhanced reward.\cite{montazeri2023scnnFinrl} In this study, a specific permutation was applied to the feature vector, thereby generating a CNN matrix that strategically positions more pertinent features in close proximity. Our comprehensive experimental evaluations unequivocally demonstrate a substantial enhancement in reward attainment.

\end{abstract}

\begin{IEEEkeywords}
CNN, DRL, Shuffled Features, Finance,
\end{IEEEkeywords}

\section{Introduction}
Deep Reinforcement Learning (DRL) has emerged as a powerful paradigm in the realm of finance, revolutionizing the way investment strategies are formulated and executed. It combines the strengths of deep learning, a branch of artificial intelligence, with reinforcement learning, a framework for learning optimal decision-making policies through trial and error. This integration allows DRL models to autonomously learn and adapt strategies in complex and dynamic financial environments.

Financial markets are characterized by intricate dynamics, influenced by an array of factors such as economic indicators, market sentiment, geopolitical events, and more. Traditional quantitative models often struggle to capture the nuances of these ever-changing conditions. DRL, however, excels in this environment by allowing agents to learn directly from data, adapting their strategies in response to evolving market conditions.

Historical market data contains a wealth of information about how assets have behaved in the past. DRL models excel at identifying intricate patterns and relationships within this data. By processing time series information, these models can discern trends, cyclical behavior, seasonal effects, and other recurring patterns that may be imperceptible to human analysts or conventional quantitative models.

One of the key advantages of DRL in finance is its ability to leverage historical market data. By training on extensive time series datasets, DRL agents can learn intricate patterns, correlations, and anomalies that may elude human analysts or conventional models. This enables them to make informed trading decisions based on a deeper understanding of market behavior.

\section{Related Works}
Recently there have been some papers published on creating financial environments,\cite{liu2022finrl,amrouni2021abides,ardon2021towards,coletta2021towards,xiong2018practical,yang2020deep,zhang2019deep} as well as applying DRL models in finance. 
Liu et al. introduce FinRL-Meta  \cite{liu2022finrl}, that is a comprehensive platform for training and evaluating data-driven reinforcement learning agents in financial markets. It provides a framework with a wide range of market environments, allowing researchers to simulate various market conditions and dynamics. FinRL-Meta offers a standardized benchmarking system, enabling the evaluation of different reinforcement learning algorithms on common financial tasks. The authors also introduce several benchmark datasets and demonstrate the effectiveness of FinRL-Meta by training agents on these datasets. The platform serves as a resource for the development and comparison of data-driven reinforcement learning techniques in the context of financial markets.

In \cite{amrouni2021abides}, Amroni et al. 
introduce ABIDES-Gym, a platform that extends OpenAI's Gym environment to facilitate multi-agent discrete event simulations. This platform is specifically designed for simulating financial markets. ABIDES-Gym provides a flexible framework for creating diverse, customizable environments that can replicate complex market dynamics and agent interactions. The authors demonstrate the versatility of ABIDES-Gym by implementing various financial market scenarios, allowing for experimentation and testing of different trading strategies and policies. The platform is expected to be a valuable tool for researchers and practitioners in finance seeking to develop, evaluate, and compare algorithms and models in a controlled, simulated environment.

Li et al. in \cite{li2021finrl} introduce a high-performance and scalable framework for applying (DRL) to quantitative finance. The authors address the computational challenges associated with training DRL models on financial data. They propose a system that optimizes distributed computing resources to accelerate training times. Additionally, the authors provide benchmark results demonstrating the efficiency and effectiveness of their approach in various financial tasks. FinRL-Podracer represents a significant advancement in the application of DRL to quantitative finance, offering improved performance and scalability for training complex models in this domain.

\subsection{PROBLEM DESCRIPTION}

\begin{figure}[htbp]
\centerline{
\includegraphics[width=0.5\textwidth]{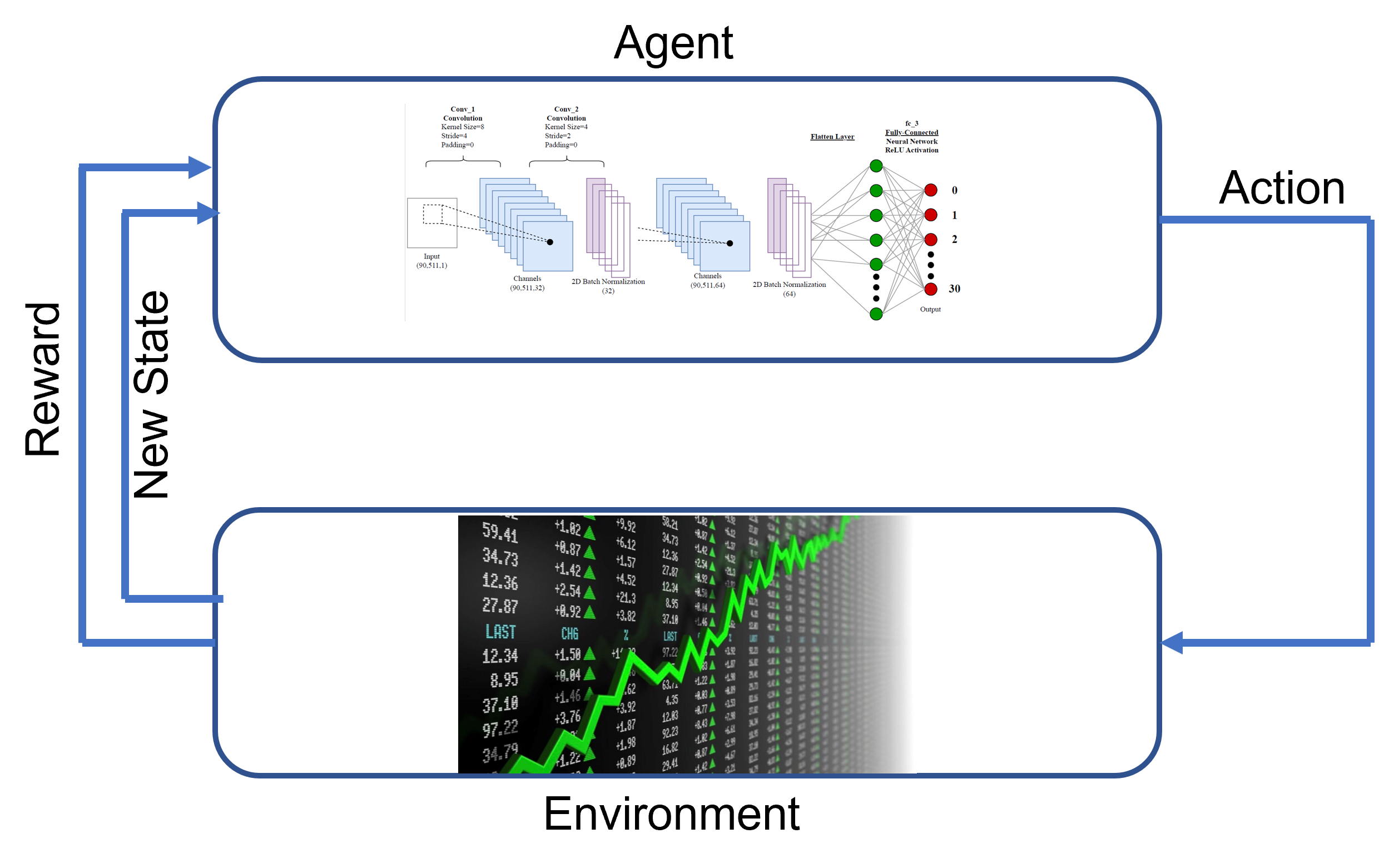}}
\caption{DRL agent and environment.}
\label{fig:drl}
\end{figure}

The behavior of a trader is to sell some stocks and buy some others. This behavior is based on stock price, fundamental data,
and other data they capture. Therefore a trader can be modeled as a Markov Decision Process (MDP) with four tuples $(S, A, r, \gamma)$, where S and A denote the state space and action space, respectively. $r(s, a, s' )$ is a reward function and $\gamma$ is the discount factor. The discount factor represents the relative importance of future rewards compared to immediate rewards. Fig. \ref{fig:drl} demonstrates the interaction of our DRL agent and our stock market environment.  

A trading agent learns a policy $\pi(s_d|a_d) $ that maximizes the discounted cumulative return shown in \eqref{eq:reward}.
\begin{equation}
     R=\ \Sigma_{d=1}^{D} \gamma^d r(s_d,a_d, s'_d) \
     \label{eq:reward}
\end{equation}
we have the following list of states, actions, rewards, policy and state action value functions to model the financial market.
\begin{itemize}
    \item State $s=[b, p, h, f]:$ a set that consist of balance $b$, price $p \in \mathbb{R}_+^D$, holdings of stock $h\in \mathbb{Z}_+^D$, and fundamental indicators $f$. where D is the number of stocks that we consider in the market. Fundamental indicators cover financial ratios.
    % listed in table \ref{tab:featurelist}.
    \item Action $a=[sell, buy, hold]:$ a set of actions for all D stocks, consisting of sell, buy, hold which leads to a reduction, growth, or no alteration in the holdings h, correspondingly.
    \item Reward $r(s, a, s'):$ The adjustment in portfolio value upon executing action "a" in state $"s"$ and transitioning to the next state $"s'"$. The portfolio value encompasses the total value of equities in the held stocks, denoted as $p^Th$, plus the remaining balance, $"b"$.
    \item Policy $\pi(s)$: The stock trading approach in state "s" entails the probability distribution of "a" in the state "s".
    \item The action-value function $Q_\pi(s, a)$
    represents the anticipated reward obtained by taking action "a" in state "s" according to policy $\pi$.
\end{itemize}

\subsection{Environment}
We have updated the environment developed by Finrl to enable the environment state representation in the form of a matrix. Our modified environment initializes the state matrix with data from the initial 90-day period of the dataset. Subsequently, we incrementally shift the matrix by one day each time a new action is taken. There are some monitor/control parameters in the environment to facilitate learning process.

In stock market modeling, turbulence refers to a period of heightened uncertainty, volatility, and instability in the financial markets. It is characterized by rapid and significant fluctuations in stock prices, which can make it difficult to model, predict, and learn to respond to market movements. 
Machine learning models typically exclude the period of uncertainty in the stock market, often characterized by a turbulence factor, in order to stabilize their learning process. In contrast, we chose not to activate the turbulence exclusion and instead closely monitored and took action during this period as well as other period of time.

The Sharpe ratio is the next parameter that measures the risk-adjusted return of an investment or a trading strategy. It was developed by Nobel laureate William F. Sharpe and is widely used in finance to assess the performance of an investment or portfolio relative to its level of risk.
The formula for the Sharpe ratio is:
\begin{equation}
\text{Sharpe Ratio} = \frac{(R_p-R_f)}{\sigma_p}
\end{equation}
Where $R_p$ is the average return of the investment or portfolio. $ R_f$ is the risk-free rate of return (the return of a risk-free investment, typically a government bond).  $\sigma_p$ is the standard deviation of the investment or portfolio's returns, which represents its volatility or risk. 
In this project, this ratio is computed based on the mean divided by the standard deviation of daily assets.

The cost of accumulated reward is another parameter that is closely monitored which corresponds to the number of sell/buy actions that are generated by the DRL agent. In practice, The cost of stock trading refers to the expenses and fees associated with buying and selling stocks in financial markets. These costs can vary depending on various factors, including the type of brokerage or platform used, the frequency of trading, the specific services offered, etc. We employ identical fixed cost percentages as those developed in the FinRL framework. 

\subsection{Feature vector}
 Feature vector encompasses the initial amount, stock price of thirty companies (table \ref{tab:comp_sectors}), number of shares held from thirty companies, and fifteen financial ratios for each one of thirty companies as it is listed in table \ref{tab:featurevectore}. 
\begin{table}[]
\centering
\caption{List of compnaies in our dataset and their sectors}
\label{tab:comp_sectors}
\begin{tabular}{|l|l|}
\hline
Sector name                                                       & Companies                                                                                                                                                                                   \\ \hline
Technology                                                        & \begin{tabular}[c]{@{}l@{}}The Home Depot Inc. (HD)\\ McDonald's Corporation (MCD)\\ Nike, Inc. (NKE) \\ The Walt Disney Company (DIS)\\ Walmart Inc. (WMT)\end{tabular}                    \\ \hline
\begin{tabular}[c]{@{}l@{}}Consumer \\ Staples\end{tabular}       & \begin{tabular}[c]{@{}l@{}}The Coca-Cola Company (KO)\\ Procter \& Gamble Co. (PG)\\ Walgreens Boots Alliance, Inc. (WBA)\end{tabular}                                                      \\ \hline
Financials                                                        & \begin{tabular}[c]{@{}l@{}}American Express Company (AXP)\\ Goldman Sachs Group, Inc. (GS)\\ JPMorgan Chase \& Co. (JPM)\\ Visa Inc. (V)\\ The Travelers Companies, Inc. (TRV)\end{tabular} \\ \hline
Healthcare                                                        & \begin{tabular}[c]{@{}l@{}}Amgen Inc. (AMGN)\\ Johnson \& Johnson (JNJ)\\ Merck \& Co.\\ Inc. (MRK)\\ UnitedHealth Group Incorporated (UNH)\end{tabular}                                    \\ \hline
\begin{tabular}[c]{@{}l@{}}Consumer\\  Discretionary\end{tabular} & \begin{tabular}[c]{@{}l@{}}The Home Depot, Inc. (HD)\\ McDonald's Corporation (MCD)\\ Nike, Inc. (NKE)\\ The Walt Disney Company (DIS)\\ Walmart Inc. (WMT)\end{tabular}                    \\ \hline
Industrials                                                       & \begin{tabular}[c]{@{}l@{}}The Boeing Company (BA)\\ Caterpillar Inc. (CAT)\\ Honeywell International Inc. (HON)\\ 3M Company (MMM)\end{tabular}                                            \\ \hline
Energy                                                            & Chevron Corporation (CVX)                                                                                                                                                                   \\ \hline
Materials                                                         & Dow Inc. (DOW)                                                                                                                                                                              \\ \hline
Telecom                                                           & Verizon Communications Inc. (VZ)                                                                                                                                                            \\ \hline
\end{tabular}
\end{table}

Financial ratios are generally derived from numerical data extracted from a company's financial statements in order to extract significant insights about its performance. The figures presented in financial statements, including the balance sheet, income statement, and cash flow statement, are employed to conduct quantitative analysis, evaluating aspects such as liquidity, leverage, growth, margins, profitability, return rates, valuation, and other relevant metrics.

 \begin{table}[htbp]
\caption{Daily Feature Vector}
\begin{center}
\begin{tabular}{|l|l|}
\hline
Name                         & size \\ \hline
Amount                       & 1    \\ \hline
Price                        & 30   \\ \hline
Share held                   & 30   \\ \hline
Financial ratios (15 * 30)     & 450  \\ \hline
Total size of feature vector & 511  \\ \hline
\end{tabular}
\label{tab:featurevectore}
\end{center}
\end{table}

\begin{table}[htbp]
\caption{Fifteen financial ratios in feature vector}
\begin{center}

\begin{tabular}{|l|l|l|}
\hline
Category & Financial ratios & Metrics                                                                                                                \\ \hline
1        & Liquidity        & \begin{tabular}[c]{@{}l@{}}Current ratio\\ Cash ratio\\ Quick ratio\end{tabular}                                       \\ \hline
2        & Leverage         & \begin{tabular}[c]{@{}l@{}}Debt ratio \\ Debt to equity\end{tabular}                                                   \\ \hline
3        & Efficiency       & \begin{tabular}[c]{@{}l@{}}Inventory turnover ratio\\ Receivables turnover ratio\\ Payable turnover ratio\end{tabular} \\ \hline
4        & Profitability    & \begin{tabular}[c]{@{}l@{}}Operating margin\\ Net profit margin\\ Return on assets\\ Return on equity\end{tabular}     \\ \hline
5        & Market value     & \begin{tabular}[c]{@{}l@{}}Earnings Per Share\\ Book Per Share\\ Dividend Per Share\end{tabular}                       \\ \hline
\end{tabular}

\label{tab:featurelist}
\end{center}
\end{table}

Table \ref{tab:featurelist} lists fifteen financial metrics that are classified into five categories liquidity metrics, leverage indicators, efficiency measures, profitability gauges, and market value assessments. Liquidity ratios assess a company's capacity to settle short- and long-term financial obligations. Leverage ratios quantify the proportion of capital sourced from debt. Essentially, these financial metrics are employed to assess a company's indebtedness. Efficiency ratios, alternatively referred to as activity financial ratios, gauge the effectiveness of a company deploying its assets and resources. Profitability ratios assess a company's capacity to generate earnings in relation to revenue, assets on the balance sheet, operational expenses, and equity. Market value ratios are employed to assess the stock price of a company.

\section{CNN ARCHITECTURE }
A CNN is a  form of neural network that is essentially designed to work on matrix state representation (2D) data sets such as images and videos. CNN is able to learn features through the optimization of filters (kernels) and it can be formed by a sequence of multiple Convolutional, Pooling, Normalization, Dropout, and Fully Connected layers. The presence of multiple consecutive layers poses training difficulties, primarily attributed to the issues of vanishing and exploding gradients. There are approaches to address the challenges of vanishing and exploding gradients encountered in earlier neural networks by employing regulated weights across fewer connections.

In the implementation of our CustomCNN class, a significant emphasis is placed on the architecture and preprocessing of the input data to adapt to the unique characteristics of financial datasets. The convolutional layers, with their varying kernel sizes and strides, are meticulously designed to capture the temporal patterns and correlations within the financial data. This is particularly crucial in the context of stock market data, where the interplay of various factors like stock prices, trading volumes, and technical indicators can be complex and highly dynamic. The use of Batch Normalization following each convolutional layer aids in stabilizing the learning process by normalizing the input to each layer, thus mitigating the internal covariate shift. This is vital in a domain like finance, where input data can exhibit significant variability and non-stationarity.

In order to employ 2D Convolutional Neural Networks (CNNs) in financial data analysis, it is imperative to restructure the representation of our environment state into a matrix format, as opposed to a vector. Financial data inherently manifests as a vector composed of daily features, wherein certain attributes like price undergo updates on a daily (or hourly, etc.) basis, while others, including revenue, income, assets, liabilities, inventories, debt, and the like, experience quarterly updates.

\begin{figure}[htbp]
\centerline
{\includegraphics[scale=0.5]{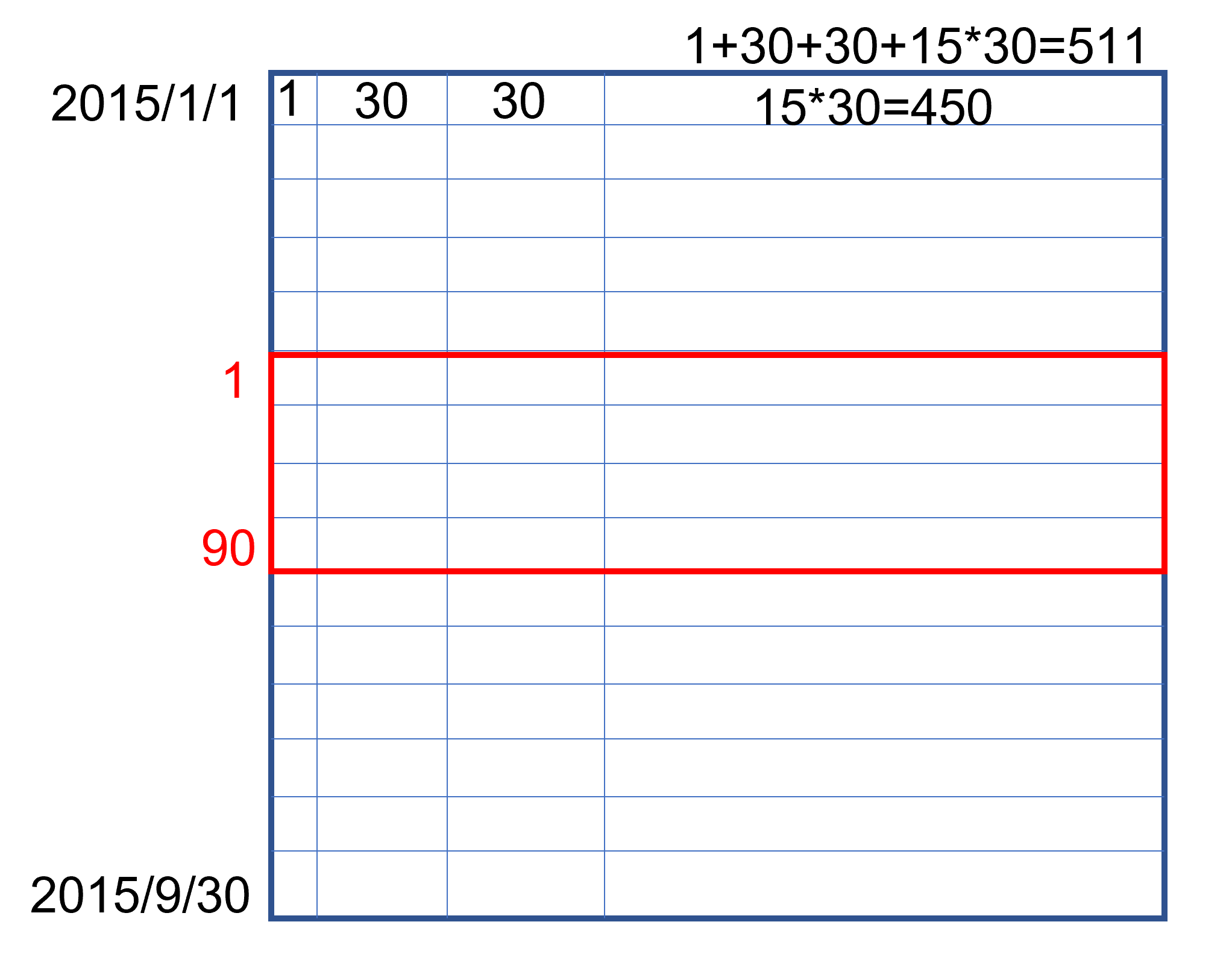}}
\caption{Sliding window to create CNN input matrix channel, before shuffling.}
\label{fig:Sliding_window}
\end{figure}

The previously designed matrix concatenates a 90-day feature vector that corresponds to a quarter to create a matrix state representation,[11] Fig. \ref{fig:Sliding_window}. The daily feature vector is the same feature vector published in Finrl \cite{liu2022finrl} which has 511 features extracted for thirty different companies.

The difference, and therefore the main idea of this research is that we have added pre-processing capabilities such that our process rearranges the input tensor to ensure that related features are positioned adjacent to each other, enhancing the network's ability to extract meaningful patterns from the data. For instance, by placing the closing prices and the number of shares for each ticker close to each other, the network can more effectively learn the relationships between these variables. This rearrangement is particularly beneficial in the context of financial data, where the relationships between different types of data (e.g., prices, volumes, technical indicators) are often more significant than the individual data points themselves. This preprocessing step, therefore, not only facilitates more efficient learning but also contributes to the robustness and generalizability of the model when applied to diverse financial datasets.

\begin{figure}[htbp]
\centerline
{\includegraphics[scale=0.5]{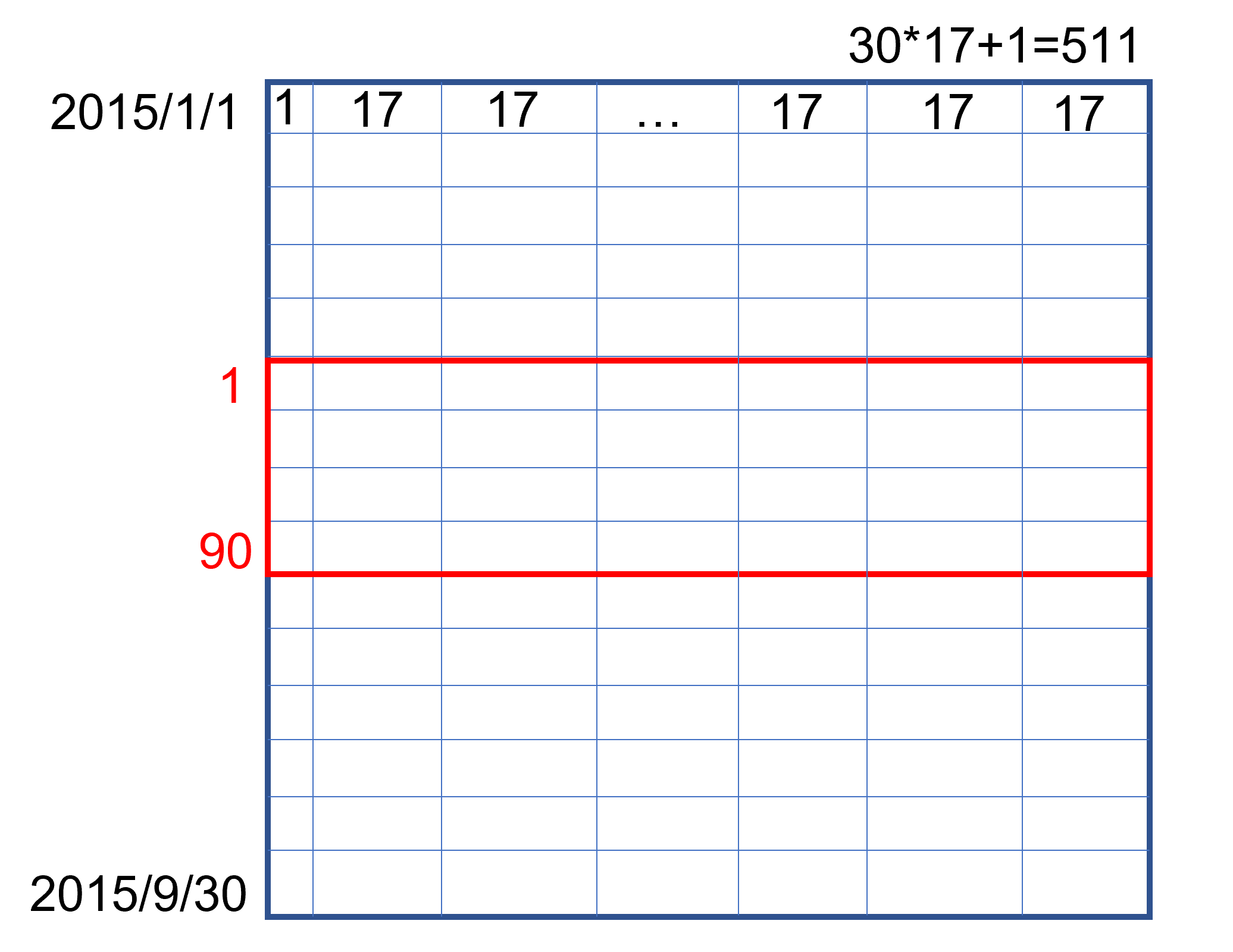}}
\caption{Sliding window to create CNN input matrix channel, after shuffling.}
\label{fig:Sliding_window_after}
\end{figure}

We applied a shuffling operation to the daily feature vector with the aim of enhancing the feature engineering process in our CNN model. Fig. \ref{fig:Sliding_window_after} illustrates our modified feature vector which is a shuffled version of the previously employed feature vector.

\begin{figure*}[htbp]
\centerline
{\includegraphics[scale=0.75]{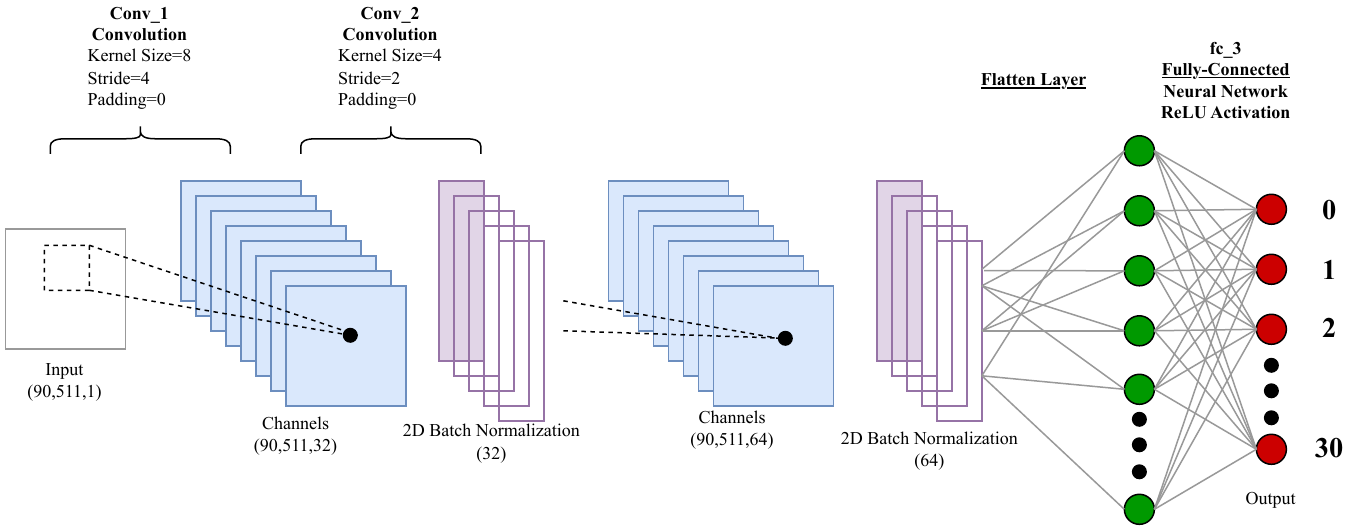}}
\caption{Our CNN architecture.}
\label{fig:CNN_arch}
\end{figure*}

After rearranging the input feature vector into a new single-channel sliding window matrix, we explored different CNN architectures. We noticed significant issues with vanishing and exploding gradients in this dataset using the CNN architecture. In response, we incorporated 2D Batch Normalization layers following the two convolutional layers (see Fig. \ref{fig:CNN_arch}). This adjustment led to a notable improvement in the DRL agent's ability to accurately perceive and adjust to environmental changes.

\section{EVALUATION}
In our study, we focus on the impact of feature vector arrangement on the performance of the custom CNN agent within the FinRL environment. The agent's architecture, as detailed in the figure, includes a convolutional neural network (CNN) that processes the input state, which consists of stock prices and technical indicators. The unique aspect of our approach lies in the rearrangement of these input features. We hypothesize that by shuffling the order of features related to different stocks and technical indicators, the CNN can potentially learn more generalized and robust representations. This is particularly relevant in the context of financial markets, where the relative importance of different stocks and indicators can vary over time. The shuffled feature vector approach aims to mitigate the risk of the model overfitting to specific patterns present in the training data, thereby enhancing its adaptability to new market conditions.

To evaluate the effectiveness of our custom CNN agent, we utilize the modified FinRL stock trading environment, introduced by Liu et al. \cite{liu2022finrl}. This framework is the first open-source platform for advancing research in financial reinforcement learning and leverages the daily state of the environment as the foundation to interact with the agent to create actions. 

These actions are expressed as vectors with a dimensionality of 30, confined within the range of (-1, +1). It's worth noting that a negative action signifies the act of selling a portion of shares, whereas a positive action involves purchasing additional shares. To quantify the scale of these transactions, the actions are further adjusted to a range of (-100, +100), representing the number of shares allocated for either sale or acquisition. The environment employs an initial reward of one million dollars, in line with established research practices. This reward is later scaled down to one when it is returned from the environment to the agent, aiding in the learning process.

We obtained the stock prices of thirty companies listed in the Dow Jones index, spanning from January 2015 to September 2023. This timeframe encompasses the pandemic, a particularly challenging period in the field of finance. This data is sourced from both Yahoo Finance and the Wharton dataset \cite{Wharton}. We subsequently selected the January 2015 to January 2023 as the training portion and left the rest of 2023 to serve as test data.

Using the dataset, we trained 3 models each with their own unique algorithms, approaches, and features. All models used the Proximal Policy Optimization \cite{ppo2017} as the managing algorithm. The first model illustrated in Fig.5 with the red line plot was an MLP agent, the second model, marked with the color blue, used our original CNN (without the feature shuffling) agent as its policy, and the third model, the green line in the plot, was trained with the custom CNN agent as its policy and had its features shuffled using the method described in Architecture section.

By even a mere glance, the shuffled-featured CNN demonstrates a clear improvement over the original CNN and decisively outperforms the MLP agent. What is most interesting is that both CNN policies, particularly the shuffled-featured one performed vastly better from January 2020 to Jan 2023, outpacing the MLP by almost 100 percent during their peak arround Oct 2023.

\begin{figure}[htbp]
\centerline{
\includegraphics[width=0.5\textwidth]{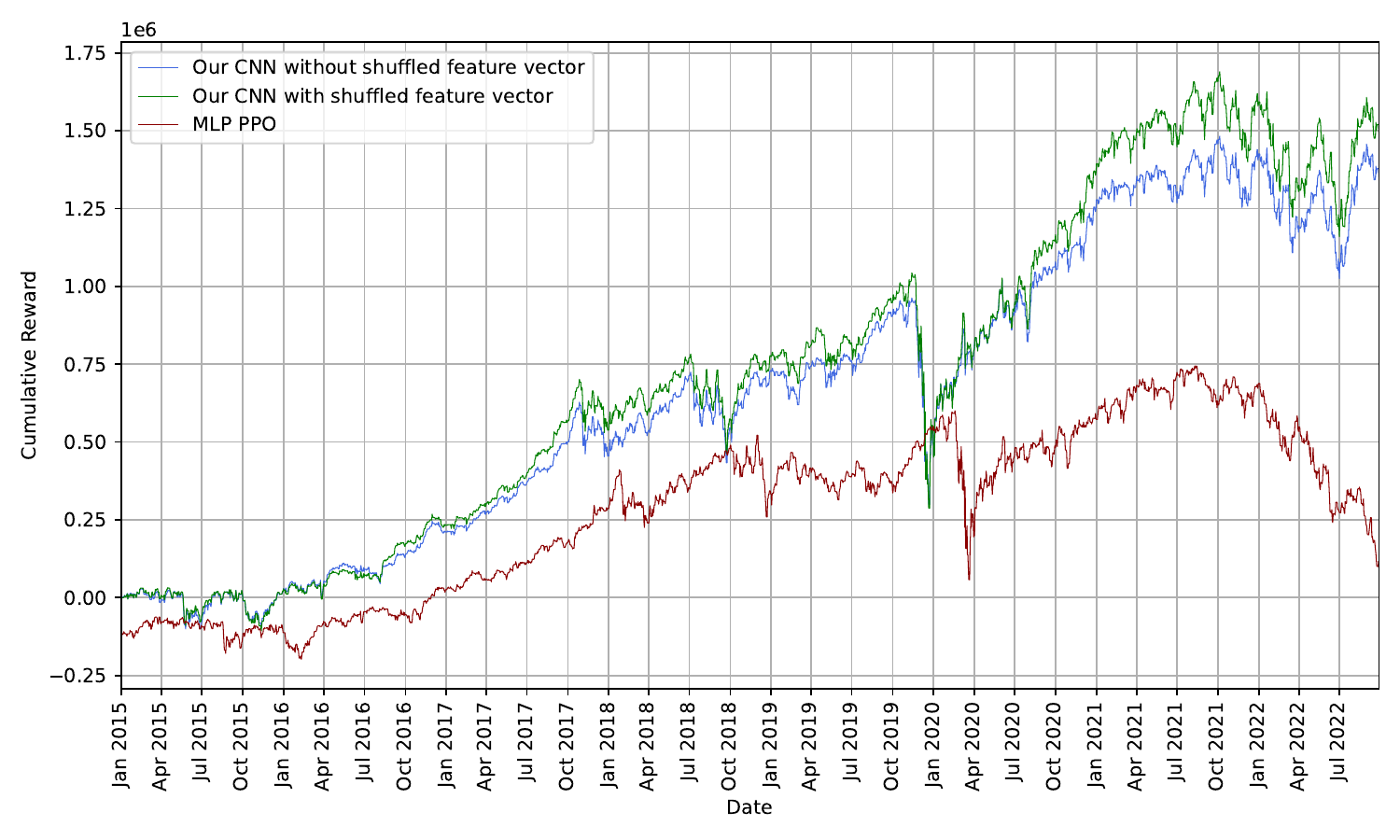}}
\caption{Training reward.}
\label{fig:training_reward}
\end{figure}

\section{Conclusion}
In the realm of financial market analysis, the robustness and adaptability of machine learning models are paramount. The implementation of a custom convolutional neural network (CNN) in this study, particularly with the inclusion of a feature vector shuffling mechanism, marks a significant stride in this direction. The architecture of the customized CNN, underscores the model's ability to efficiently process and learn from multi-dimensional data. This is evident in its handling of varying stock dimensions and technical indicators, a critical aspect in the dynamic and often unpredictable domain of stock trading. The model's design, featuring layers of convolutional and batch normalization followed by ReLU activation, is tailored to extract and learn intricate patterns from financial datasets. This approach, when juxtaposed with traditional MLP models, demonstrates a superior capacity in capturing the nuances of market data, thereby enhancing prediction accuracy and reliability.

This study showcases the capability of convolutional neural networks as a proficient deep reinforcement learning framework for stock trading in settings characterized by shuffled feature vectors. Through empirical assessments using historical market data and the FinRL platform, our specialized CNN agent in addition to our permutation of feature vector demonstrated consistent learning and elevated cumulative rewards. In contrast, conventional MLP models exhibited significant declines in performance at the beginning.

Our assessments show the intrinsic capability of CNNs with the shuffled feature vector to derive significant features from financial data. Future studies can extend these findings by examining the adaptability of this approach to different assets, markets, and datasets with higher-frequency trading data.

Furthermore, the empirical evaluation of the model's performance, as illustrated in the paper, offers a comprehensive visualization of the learning process over time. We provided a clear and quantifiable measure of the model's efficacy which served not only as a testament to the model's proficiency in handling shuffled feature vectors but also layed the groundwork for future enhancements. The potential for integrating this CNN architecture with more complex and varied datasets, including high-frequency trading data, opens new avenues for research. Such explorations could delve into the model's scalability and effectiveness across different market conditions and asset classes. This would not only validate the model's versatility but also contribute significantly to the development of more sophisticated and robust trading algorithms in the field of quantitative finance.
\bibliographystyle{IEEEtran} % We choose the "plain" reference style
\bibliography{biblio} 

% \bibitem{b1} G. Eason, B. Noble, and I. N. Sneddon, ``On certain integrals of Lipschitz-Hankel type involving products of Bessel functions,'' Phil. Trans. Roy. Soc. London, vol. A247, pp. 529--551, April 1955.
% \bibitem{b2} J. Clerk Maxwell, A Treatise on Electricity and Magnetism, 3rd ed., vol. 2. Oxford: Clarendon, 1892, pp.68--73.
% \bibitem{b3} I. S. Jacobs and C. P. Bean, ``Fine particles, thin films and exchange anisotropy,'' in Magnetism, vol. III, G. T. Rado and H. Suhl, Eds. New York: Academic, 1963, pp. 271--350.
% \bibitem{b4} K. Elissa, ``Title of paper if known,'' unpublished.
% \bibitem{b5} R. Nicole, ``Title of paper with only first word capitalized,'' J. Name Stand. Abbrev., in press.
% \bibitem{b6} Y. Yorozu, M. Hirano, K. Oka, and Y. Tagawa, ``Electron spectroscopy studies on magneto-optical media and plastic substrate interface,'' IEEE Transl. J. Magn. Japan, vol. 2, pp. 740--741, August 1987 [Digests 9th Annual Conf. Magnetics Japan, p. 301, 1982].
% \bibitem{b7} M. Young, The Technical Writer's Handbook. Mill Valley, CA: University Science, 1989.
% \end{thebibliography}
% \vspace{12pt}
% \color{red}
% IEEE conference templates contain guidance text for composing and formatting conference papers. Please ensure that all template text is removed from your conference paper prior to submission to the conference. Failure to remove the template text from your paper may result in your paper not being published.

\end{document}